\documentclass[a4paper,12pt]{article} 

\usepackage{a4wide,amsmath,latexsym,amsthm,amssymb,pictex,graphicx}

\providecommand{\text}{\mbox}
\providecommand{\eqref}[1]{(\ref{#1})}

\def\t{\tau}

\newcommand{\real}{{\mathbb R}} 
\newcommand{\N}{{\mathbb N}}

\newcommand{\integer}{{\mathbb Z}} 
 
\newcommand{\half}{\frac{1}{2}} 
\def\aff{\mathop{\rm aff}}

\begin{document} 
\title{New group structures for Carbon onions and\\ Carbon nanotubes via\\
affine extensions of non-crystallographic \\Coxeter groups}
\author{R. Twarock\\ Department of Mathematics, City University \\ Northampton Square, London EC1V 0HB\\
r.twarock@city.ac.uk}
\maketitle 

\begin{abstract} 
We present results underlining the conjecture \cite{Twarock:2002} that 
the affine extensions for non-crystallographic Coxeter groups introduced in this reference are suitable mathematical objects for the description of the symmetries of Carbon onions and Carbon nanotubes. 
It is the hope that these considerations will shed new light on open questions concerning structure, stability and formation of these fullerenes. 
\end{abstract}

\section{Introduction}\label{s1} 

Group theory is a powerful tool in crystallography, and helps to understand issues like structure, formation and stability of crystallographic objects. It is the purpose of this paper to provide evidence for the conjecture that a new symmetry structure recently introduced in \cite{Twarock:2002} appears to be a suitable object for a mathematical description of particular types of fullerenes such as 
Carbon onions and Carbon nanotubes. 

Carbon onions consist of nested cages of Carbon atoms \cite{Kroto:1992}. From a group theoretical point of view, the individual cages are characterized by their symmetry properties which are given by the non-crystallographic Coxeter group $H_3$ \cite{Humphreys:1992}, which also describes the buckminster fullerene $C_{60}$ \cite{Kroto:1985}. A group theoretical description of the symmetry of a Carbon onion as a whole, however, is lacking at present, but is needed in order to understand the peculiarities of these objects. 
We demonstrate here that the technique of affine extensions of non-crystallographic Coxeter groups, which was developed in  \cite{Twarock:2002} for the description of fragments of aperiodic point sets or quasicrystals, leads in the case of $H_3$ to an affine extension $H_3^{\aff}$ which may describe the symmetries of Carbon onions as entities. In particular, we show that the generators of $H_3^{\aff}$ may be used to obtain three dimensional point sets from a seed configuration which are arranged concentrically around the origin and bear the characteristic shell structure of Carbon onions. We show numerical studies based on $H_3^{\aff}$ generated models and compare the results with data about Carbon onions \cite{Kroto:1992}. This comparison suggests to understand the different shells of a Carbon onion as different $H_3$ orbits, connected by the $H_3^{\aff}$ symmetry via the action of the further group generator which is introduced for the affine extension. 

The way this mathematical models for Carbon onions is built from the group action of $H_3^{\aff}$ allows for flexibility, and we point out a further generic possibility here: If the number of generators in $H_3^{\aff}$ is restricted from four to three in a suitable way, one obtains along similar lines as for Carbon onion models extended tubular structures. 
These bear the characteristics of Carbon nanotubes, that is extended Carbon structures with a preferred direction of growth \cite{Iijima:1992,Kroto:1997}. We give several arguments for how the approach based on $H_3^{\aff}$ symmetry explains characteristic features of Carbon nanotubes which are pointed out for example in \cite{Bonard:1998} or the references therein. 

It is possible that this study will shed new light on the mechanism of formation of Carbon onions and Carbon nanotubes, and thus will complement existing theories (e.g. \cite{Kroto:1988}). Furthermore, the new symmetries may contribute to the discussion of stability issues as discussed in the conclusion.   
\medskip

\noindent
The paper is organized as follows: 

After a short introduction of the non-crystallographic Coxeter group $H_3$ and a short description of the technique of affine extensions, presented for the example of $H_3$, we point out two ways of deriving three dimensional point sets -- and thus models for the location of atoms -- from the group action of the generators of the affine extension $H_3^{\aff}$ of $H_3$.  One method is along the lines of \cite{Twarock:2002} and leads to Carbon onion structures, the other one is a modification thereof and corresponds to Carbon nanotube structures. 
Numerical computations for these models are presented and various planar subsets are depicted. The results for 
a two dimensional cut along a fivefold axis of one of these three dimensional models is compared with the electron microscope simulations for fullerenes presented in \cite{Kroto:1992} to underline the good fit of the models with fullerene data. 
In a next step, we discuss the modifications necessary to accommodate Carbon nanotubes and discuss how specific properties of Carbon nanotubes are reflected by the model. 

Finally, we discuss open questions arising from this study, pointing to future research along these lines.

\section{Group Theoretical Building Blocks}\label{s2} 

\subsection{The non-crystallographic Coxeter group $H_3$}\label{s2.1} 

Coxeter groups are reflection groups, comprising also the Weyl groups, which are related to the classical crystallographic setting, semisimple Lie algebras and Lie groups. In three dimensions, which is the setting relevant for the study of real life objects such as fullerenes, there exists only one non-crystallographic Coxeter group, $H_3$. 
It will be discussed in detail here because all further considerations are based on it.   

Coxeter groups are characterized by a root system, that is a finite collection $\Delta$ of nonzero vectors in Euclidean space $\mathbb{E}$, satisfying 
\begin{enumerate}
\item $\Delta \cap \real \alpha = \{ \alpha, -\alpha \}\,,\qquad \forall \alpha \in \Delta$
\item $r_\alpha \Delta = \Delta\,, \qquad \forall \alpha \in \Delta$ 
\end{enumerate}
where $r_\alpha$ is given by 
\begin{equation}\label{Weyl} 
r_\alpha v=v-\left(\frac{2(v|\alpha)}{(\alpha|\alpha)}\right)\alpha\,,\qquad \alpha \in \mathbb{E}\,.
\end{equation} 
and $(.|.)$ denotes the inner product in $\mathbb{E}$. 

(\ref{Weyl}) shows that a Coxeter group can be specified by indicating a root system for it. 
For $H_3$, the root system consists of 30 roots and can be modeled as \cite{CKPS} 
\begin{equation}\label{icosH3}
\Delta_3 = \left\lbrace{
\begin{array}{cl}
(\pm 1,0,0) & \mbox{ and all permutations }\\
\half(\pm 1,\pm \t',\pm \t) & \mbox{ and all even permutations }
\end{array}
}\right\rbrace\,
\end{equation}
where $\tau$ and $\tau'$ are irrational numbers given as 
\begin{equation}\label{tau}
\begin{array}{rcl}
\t & := & \half(1+\sqrt{5}) \\
\t' & := & \half(1-\sqrt{5})=1-\t=-\frac1\t\,.
\end{array}
\end{equation}
Geometrically, the root polytope of  $H_3$ is formed by 12 equilateral pentagons and 20 equilateral triangles. It has 30 vertices given by the elements in $\Delta_3$ and 60 edges. Alternatively, one may view the roots of
$\Delta_3$ given in \eqref{icosH3} as icosians \cite{Moody:1993,CMP}, that is purely imaginary
quaternions of special kind.

The root system of $H_3$ can be expressed in term of a subset, called basis of simple roots. In the orthonormal basis, it is given as 
\begin{equation}
\alpha_1=(0,0,1)\,,\quad\alpha_2=\tfrac12(-\t',-\t,-1)\,,\quad
\alpha_3=(0,1,0)\,.
\end{equation}
The information about the simple roots can be encoded in the Cartan matrix: 
 \begin{equation}
A:=(a_{ij})=
\left(\frac{2(\alpha_i\mid\alpha_j)}{(\alpha_j|\alpha_j)}\right)
= ((\alpha_j|\alpha_k)) 
=
\begin{pmatrix}
 2 &    -1 & 0 \cr 
-1 &    2  & -\t \cr
 0 & -\t &   2  
\end{pmatrix}
\end{equation}
According to (\ref{Weyl}) this encodes the generators of the group, and allows to derive an explicit representation for them using this formula. 
Furthermore, the group relations can be read off from the Cartan matrix as follows, using $r_k \equiv r_{\alpha_k}$ to shorten notation: 
\begin{equation}\label{2reflect}
(r_jr_k)^M= 1 \quad\mbox{ where }\quad
\left\{\begin{matrix}   
M=1 \quad &\mbox{ if }\ &a_{jk}&= 2\\
M=2 \quad &\mbox{ if }\ &a_{jk}&= 0\\
M=3 \quad &\mbox{ if }\ &a_{jk}&=-1\\
M=5 \quad &\mbox{ if }\ &a_{jk}&=-\t
\end{matrix}\right.
\end{equation}

Objects with $H_3$ symmetry have correspondingly $10$-, $6$- and $4$-fold rotational symmetry axes and from (\ref{2reflect}) it is clear that the generators corresponding to these symmetries are obtained via a restriction of the group generators to a corresponding subgroup. For instance, a 10-fold symmetry axis corresponds to the action of the subgroup $H_2$ of $H_3$, given by the action of the generators $r_2$ and $r_3$ in (\ref{2reflect}).  
 
\subsection{$H_3^{\aff}$ as an affine extension of $H_3$}\label{s2.2} 

Affine extensions of non-crystallographic Coxeter groups have been introduced and discussed in \cite{Twarock:2002}, and we thus only briefly indicate the main steps here. 

In general terms, the idea is to introduce a further group generator. Due to the correspondence between Cartan matrix and group generators pointed out above, this can be done via an extension of the Cartan matrix. 
This process is subject to conditions imposed on the extended Cartan matrix $A=(a_{ij})$, and in the non-crystallographic case these are 
\begin{equation}\label{reqII}
a_{ii}=2\,,\quad
a_{ij}=a_{ji}\,,\quad
a_{ij}\in \integer[\t]^-:=\lbrace
     x\in\integer[\t]\mid x\leq0\rbrace\,,\quad
\det(a_{ij})=0\,,
\end{equation}
where 
$\integer[\t]:=\lbrace
a+\t b\mid a,b\in\integer\rbrace$ with $\t$ as in (\ref{tau}). 

In particular, in the case of $H_3$, this leads to the following unique (see \cite{Twarock:2002}) extension of the Cartan matrix of $H_3$ 
\begin{equation}\label{Cartan}
\begin{pmatrix}
 2    &  0  & \t' & 0     \\ 
 0    &  2  &  -1   & 0     \\
\t' &  -1 &   2   & -\t \\
 0    &  0  & -\t &   2  
\end{pmatrix}\,,
\end{equation}
with $\t$, $\t'$ as in (\ref{tau}),  
where the first line or column contain information about the new generator. 
Then according to (\ref{Weyl}) one obtains representations for the four generators of $H_3^{\aff}$. We indicate their  action on a vector $v=(v_1,v_2,v_3)$ which we consider for convenience with coordinates in the basis of fundamental weights (or $\omega$-basis)  $\{\omega_1,\omega_2,\omega_3\}$, which is the basis dual to the basis of simple roots $\{\alpha_1,\alpha_2,\alpha_3\}$ and which is given by    
\begin{equation}
(\alpha_j|\omega_k)=\tfrac12(\alpha_j|\alpha_j)\delta_{jk}
=\delta_{jk}\,.
\end{equation}
One obtains 
\begin{equation}\label{trafos3}
\begin{aligned}
T v   & = v+\alpha_H &&=(v_1,v_2-\t', v_3)\\
r_1 v & = v-2(v|\alpha_1)\alpha_1  &&=(-v_1, v_1+v_2,v_3)\\
r_2 v & = v-2(v|\alpha_2)\alpha_2  &&= (v_1+v_2, -v_2, v_3+\t v_2)\\ 
r_3 v & = v-2(v|\alpha_3)\alpha_3  &&= (v_1, v_2+\t v_3,-v_3)\,.
\end{aligned}
\end{equation}
\bigskip
We remark that the action of $T$ corresponds to a translation by the highest root vector 
$\alpha_H= \t\alpha_1+2\t\alpha_2+\t^2\alpha_3= -\t'\omega_2$, which in cartesian coordinates corresponds to $(1,0,0)$, that is the translation is along one Cartesian coordinate direction. 

Note that in order to obtain the group relations satisfied by (\ref{trafos3}), (\ref{2reflect}) has to be extended by 
$M=5 \quad \mbox{ if }\ a_{jk} =\t'$.

\section{Models for Atomic Configurations based on $H_3^{\aff}$}\label{s3} 

There are several options to construct three dimensional point sets with $H_3^{\aff}$ symmetry as models for atomic configurations in fullerenes. 
The simplest possibility is to consider point sets which are obtained from a seed configuration under an iterate application of the group generators of $H_3^{\aff}$. Clearly, due to the fact that $H_3$ is non-crystallographic, the unrestricted action of the full group leads to a dense filling of $\real^3$, and one needs to apply appropriate restrictions. 
\medskip

We are considering two types of restrictions in this letter: 

\begin{enumerate}
\item A restriction of the number of times the generator of the central extension, $T$, is allowed to act on the initial configuration. 
\item A restriction of the four generators of $H_3^{\aff}$ to a subset of three generators containing $T$ and two further reflections. 
\end{enumerate}

Both restrict the group action of $H_3^{\aff}$ by restricting the monomials built from the group generators of $H_3^{\aff}$ to an allowed subset. In particular, we have  

\begin{enumerate}
\item monomials with a limited occurrence  of the generator $T$. Since the actions of the other generators are cyclic, this restriction leads to a finite set for the subset of allowed monomials. 
\item monomials with a (possibly) infinite occurrence of $T$, but not containing one of the other three (cyclic) group generators. If in addition the occurrence of $T$ is restricted, finite tubular structures are modeled, otherwise, the structures are generically infinite. 
 \end{enumerate}

The first case corresponds  to Carbon onions and the second one to Carbon nanotubes, and we discuss these cases separately below.

\subsection{Carbon onions}\label{s3.1} 

This case is similar to the models for $H_2^{\aff}$-induced quasicrystals introduced and discussed in \cite{Twarock:2002}. 
Let $s^m(T,r_1,r_2,r_3)$ denote the set of all sequences formed by the
operators $T$ and $r_1,\ldots,r_3$ in which $T$ appears precisely $m$
times, that is all monomials formed by generators of $H_3^{\aff}$ such that $T$ appears precisely $m$ times, and let O denote the origin of coordinates. Then the point sets are given by the action of these monomials on the seed point O, that is   
\begin{equation}\label{affQC3}
 Q_3(n):=\lbrace s^m(T,r_1,r_2,r_3)\, O \mid m\leq n\rbrace
\end{equation} 
$n$ is called the cut-off-parameter. 
Note that $Q_3(n)$ is a family of $3$-dimensional point sets depending on $n$.  
These point sets are thus characterized by 
\begin{enumerate}
\item a choice of an initial configuration, chosen to be the origin O. 
\item a cut-off parameter $n$, which restricts the frequency of the action of $T$, and thus the monomials formed by the generators of $H_3^{\aff}$. 
\end{enumerate}

Note that there is some freedom in the choice of the seed configuration. However, O is a canonical choice since after one iteration step, the root system of $H_3$ is obtained, that is $Q_3(1) = \Delta_3$, and thus $H_3$ symmetry -- which is characteristic of individual fullerene shells -- is automatically obtained. Furthermore, any choices of seed configurations coinciding with a subset of $\Delta_3$ necessarily lead to the same result, because due to the action of $H_3$ the full root system $\Delta_3$ is reproduced in the first iteration step.

Analogously to Proposition 6.1 in \cite{Twarock:2002} it follows that from a geometric point of view, the point set obtained under the action of the allowed monomials subject to a cut-off parameter $n$ on the seed configuration  in (\ref{affQC3}) corresponds to the point set obtained via arbitrary linear combinations of up to $n$ vectors from $\Delta_3$, that is linear combinations of up to $n$ (not necessarily different) icosians: 
\begin{equation}
Q_3(n)=\left\lbrace \sum_{j=0}^{29} n_j\xi_j \mid \xi_j \in \Delta_3, 
n_j\in \N^0, \sum_{j=0}^{29} n_j=l\leq n \right\rbrace\,.
\end{equation}

Using this expression, the three dimensional point sets $Q_3(n)$ can be computed explicitly for a given cut-off parameter $n$. In figures 1 and 2, we display two-dimensional cuts which contain the origin and are perpendicular to a $6$- and $4$-fold rotational symmetry axis for the cut-off parameter 3. 

\begin{figure}
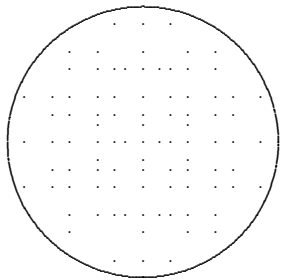

\centerline{\beginpicture
 \setcoordinatesystem units <0.600cm,0.600cm>
 \setplotarea x from -4 to 4, y from -4 to 4
 \multiput {\tiny{.}} at "full2-3.pic"
 \circulararc 360 degrees from 3.000 0 center at 0 0 
\endpicture}
\caption{Cut perpendicular to a 4-fold axis through the origin with cut-off $n=3$.}
\end{figure}

\begin{figure}
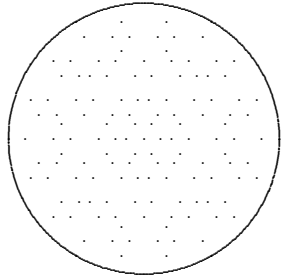

\centerline{\beginpicture
 \setcoordinatesystem units <0.600cm,0.600cm>
 \setplotarea x from -4 to 4, y from -4 to 4
 \multiput {\tiny{.}} at "full3-3b.pic"
 \circulararc 360 degrees from 3.000 0 center at 0 0 
\endpicture}
\caption{Cut perpendicular to a 6-fold axis through the origin with cut-off $n=3$.}
\end{figure}

Furthermore, we show the effect of the growth parameter on the plane corresponding to 10-fold symmetry in the figures 4 -- 6. 

\begin{figure}
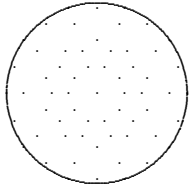

\centerline{\beginpicture
 \setcoordinatesystem units <0.600cm,0.600cm>
 \setplotarea x from -4 to 4, y from -4 to 4
 \multiput {\tiny{.}} at "full5-2.pic"
 \circulararc 360 degrees from 2.000 0 center at 0 0 
\endpicture}
\caption{Cut perpendicular to a 10-fold axis through the origin with cut-off $n=2$.}
\end{figure}

\begin{figure}
\centerline{\beginpicture
 \setcoordinatesystem units <0.600cm,0.600cm>
 \setplotarea x from -4 to 4, y from -4 to 4
 \multiput {\tiny{.}} at "full5-3.pic"
 \circulararc 360 degrees from 3.000 0 center at 0 0 
\endpicture}
\caption{Cut perpendicular to a 10-fold axis through the origin with cut-off $n=3$.}
\end{figure}

\begin{figure}
\centerline{\beginpicture
 \setcoordinatesystem units <0.600cm,0.600cm>
 \setplotarea x from -4 to 4, y from -4 to 4
 \multiput {\tiny{.}} at "full5-4.pic"
 \circulararc 360 degrees from 4.000 0 center at 0 0 
\endpicture}
\caption{Cut perpendicular to a 10-fold axis through the origin with cut-off $n=4$.}
\end{figure}

\begin{figure}
\centerline{\beginpicture
 \setcoordinatesystem units <0.600cm,0.600cm>
 \setplotarea x from -4 to 4, y from -4 to 4
 \multiput {\tiny{.}} at "full5-5.pic"
\circulararc 360 degrees from 5.000 0 center at 0 0 
\endpicture}
\caption{Cut perpendicular to a 10-fold axis through the origin with cut-off $n=5$.}
\end{figure}

To compare the configurations of point sets obtained above with data about fullerenes, we compare with the election-microscope simulation of fullerene concentric shells presented in \cite{Kroto:1992}, which is given in figure 7.  

\begin{figure}[ht]
\begin{center}
\includegraphics[width=4.5cm,keepaspectratio]{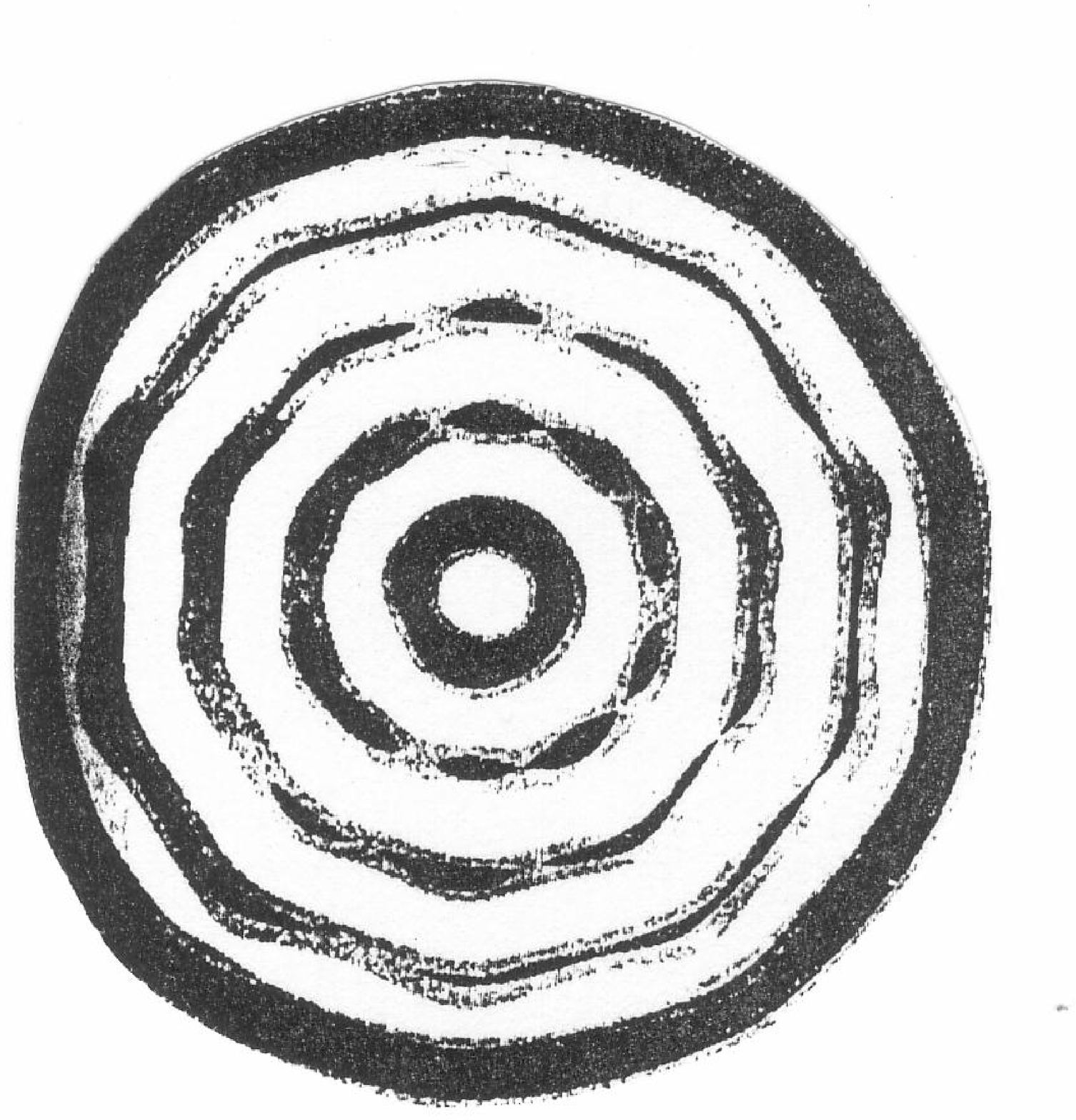}\\*
\end{center}
\caption{Election-microscope simulation of fullerene concentric shells \cite{Kroto:1992}}
\end{figure}

We observe that this picture has a very good qualitative fit with the point set in figure 4 corresponding to the $10$-fold rotational axis with cut-off parameter 3. To facilitate comparison, we depict the latter once more in figure 8, where we highlight the similarities by connecting points which are located on the same $H_3$ orbit, and by filling gaps alternatingly in black and white\footnote{Black fillings may contain further orbit, compare with figure 4.}.

\begin{figure}[ht]
\begin{center}
\includegraphics[width=4.5cm,keepaspectratio]{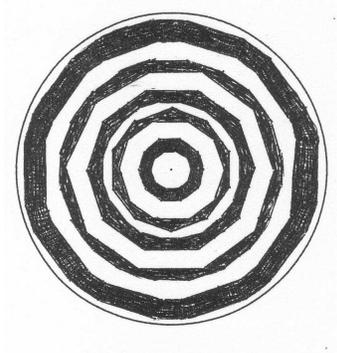}\\*
\end{center}
\caption{The model based on $H_3^{\aff}$ with cut-off parameter 3 -- cut perpendicular to a $10$-fold rotational axis.}
\end{figure}

This correspondence between figure 7 and figure 8 can be used to explain among others the peculiar shape of the second inner circle in figure 7: mathematically, it stems from the fact that the corresponding $H_3$ orbits in the mathematical picture are shifted with respect to each other, and the amount of the shift is dictated by the underlying $H_3^{\aff}$ symmetry, thus not visible in considerations based merely on $H_3$.

\subsection{Carbon nanotubes}\label{s3.2} 

Carbon nanotubes are tubular graphitic structures, which may be nested or appear as a single tube. Experimental data show that they occur with a wide distribution of lengths and diameters. Observations show furthermore \cite{Bonard:1998} that even if deformed under external forces, nanotubes regain their initial form without any apparent damage, which points to the fact that there must be a strong organizing principle underlying their structure, and we suggest here the group structure of $H_3^{\aff}$ as a possible explanation. 

From the point of view taken in this paper, nanotube structures arise naturally via a restriction of the generators of $H_3^{\aff}$: if instead of the full $H_3^{\aff}$ symmetry only group elements obtained from two of the three generators $r_k$, $k=1,2,3$, together with $T$, are admitted, then we obtain tubular structures, where the preferred direction of growth coincides with the translation direction of $T$. It is important to note that we do not mean here a restriction of the Cartan matrix (\ref{Cartan}) of $H_3^{\aff}$ to a $3\times 3$ sub-matrix ${\tilde A}$ obtained via deleting a row and a corresponding column, which would necessarily lead to a finite dimensional group since then $\det {\tilde A} \not=0$; we rather mean a restriction of the monomials formed by generators from $H_3^{\aff}$ to a subset which is such that it does not contain one of the cyclic generators $r_k$, $k=1,2,3$, but contains all possible monomials formed by all other generators, that is two cyclic generators $r_k$ and the generator of the affine extension, $T$. 
Note also that such a restricted $H_3^{\aff}$ symmetry is still represented as a three-dimensional object, and thus does not coincide with an $H_2^{\aff}$ symmetry (for an extensive discussion of this symmetry, see \cite{Twarock:2002}), which is confined to a 2 dimensional setting. In particular, this means that the translation direction of the generators, which are introduced via the affine extension in both cases, differ, and while the translation direction in the $H_2^{\aff}$ setting is given by the highest root of $H_2$ and thus is collinear with the plane orthogonal to the rotation axis, the direction of translation of the corresponding generator in $H_3^{\aff}$ does not lie in the plane corresponding to the $H_2$ (sub-)symmetry of $H_3$. Thus, a restriction of $H_3$ to its subgroup $H_2$ and a consequent affine extension of the latter to $H_2^{\aff}$ leads to a different result than selecting in the affine extension of $H_3^{\aff}$ particular monomials by restricting the number of generators to a subset. This is depicted in the following diagramme: 

\beginpicture
 \setshadegrid span <.5truept>
 \setcoordinatesystem units <1cm,1cm> point at 0 0
 \setplotarea x from -4.5 to 4.5, y from -1.4 to 4.4

 \plot 1.8 0  -1.8 0 /
 \plot -2.8 0.4  -2.8 2.6 /
 \plot  1.8 3.0  -1.8 3.0 /

\put {$>$} at 1.8 0
\put {$>$} at 1.8 3.0 

\setdashes <1mm> \setlinear 
\plot 2.8 0.4  2.8 2.6 /

	 \put {$H_3$} at -2.8 0 
   \put {$H_3^{\aff}$} at 2.8 0
	 \put {$H_2$} at -2.8 3.0 
   \put {$H_2^{\aff}$} at 2.8 3.0

\put {\scriptsize affinization} at 0 2.6
\put {\scriptsize restriction} at -4.0 1.7
\put {\scriptsize to subgroup} at -4.0 1.4
\put {\scriptsize not possible by} at 4.0 1.7
\put {\scriptsize restriction of } at 4.0 1.4
\put {\scriptsize generators} at 4.0 1.1
\put {\scriptsize affinization} at 0 -0.3
 \endpicture


This shows that in the context of tubal three dimensional structures such as Carbon nanotubes, only affine extensions of three dimensional group structures such as $H_3$ can be useful. Since $H_3$ is the only non-crystallographic Coxeter group in three dimensions -- as opposed to the existence of infinitely many non-crystallographic Coxeter groups in two dimensions --  the method for deriving three dimensional structures from a group theoretical approach based on affine extensions of non-crystallographic Coxeter groups presented here is exhaustive and $H_3^{\aff}$ symmetry plays a distinct role. Variations can only come from modifications of the seed configuration, cut-off parameter, or selection rules defining the admissible subsets of monomials. 
  
In order to obtain multishell structures of nanotube type, one has to use planar initial configurations composed of several connected, nested sets, for example circles of different radii, which are located in the plane through the origin and orthogonal to the rotation axis corresponding to the two generators $r_i$, $r_j$, $i$, $j\in \{1,2,3\}$ defining the restriction of $H_3^{\aff}$. Then, the whole setting is propagated by the generator of the affine extension $T$. 

In this picture, it is possible to explain certain phenomena which have been observed experimentally in Carbon nanotubes: 
\begin{enumerate}
\item In most cases, the layers of Carbon nanotubes have helicities, that is the carbon bonds form a spiral around the  cylinder \cite{Iijima:1992}. This spiralling effect may be explained in the above models by the action of the generators corresponding to the restriction of $H_3^{\aff}$. For instance, this may be the generators $r_2$ and $r_3$ in (\ref{trafos3}), which models rotation about a 10-fold axis according to (\ref{Cartan}) applied to (\ref{2reflect}). The combined action  of the translation along a preferred axis, given by $T$, and a rotation about a 10-fold axis (not given by the direction of $T$) necessarily leads to the spiralling effect observed in Carbon nanotubes. 

\item Another observation is that in multishell nanotubes the different shells assist each other during the growth \cite{Gamaly:1995,Charlier:1997}. This may be explained in the framework of models based on $H_3^{\aff}$ by the fact that starting with any planar initial configuration composed of disjoint objects, nested set of circles say, these configurations  are all acted upon by the same group operations and hence their evolution in three-dimensional space under the action of the group generators is necessarily correlated. 
\end{enumerate}

\section{Conclusions}\label{s4} 

The models presented here are the most simple conceivable models based on an $H_3^{\aff}$ symmetry: In the first case, all group generators of $H_3^{\aff}$ act on the initial configuration, which leads to arrangements of concentric shells typical for Carbon onions. In the second case, the number of $H_3^{\aff}$ generators is restricted such that only three of the four generators are acting, which leads to the typical tube structure of Carbon nanotubes. 

Though these are canonical choices to model structures with an $H_3^{\aff}$ symmetry, there is flexibility in these models by changing the growth parameter or by starting from different seed configurations. The choice adapted here is a natural one from the mathematical point of view: we choose the seed point O which leads to the root system of $H_3$ in the first iteration step for the case of an action of the full set of generators as explained above, and the root system (and rescalings thereof in the multilayer case) of the subgroup of $H_3$ corresponding to the restriction of $H_3^{\aff}$ in the restricted case. An analysis of concrete experimental data may give evidence for how this freedom may be chosen to meet the individual settings. For instance, the growth parameter governs the number of points in the model, that is the number of atoms in the sample, and indications for a suitable choice of this parameter  may come from a comparison with the cage structures of the shells and their number of atoms in concrete samples. These flexibilities should be exhausted in dialog with chemist experts in the field.

Another possibility to generalize these models is to introduce statistical weights on the frequency of the action of the group generators of $H_3^{\aff}$ on the initial configuration, such that the relative frequency of  appearance of the generators in the monomials is distinct, thus leading to a different selection rule for admissible subsets of monomials.  Geometrically, this will result in a breaking of the pure shell structures for Carbon onions, or a breaking of the concentric cylinders in the Carbon nanotube case. Experimental evidence points to the fact that nested structures are preferred configurations, with the scroll structure appearing only if defects are present in a tube \cite{Tsang:1996,Ebbesen:1993,Amelincx:1995}. Thus, defect structures may be explained by a deviation from the 
ideal mathematical situation of equal frequency of occurrence. It is quite possible that such considerations are related to questions of stability and energy balance in the sense that certain statistical weights are preferred from this point of view. 

In any case, the fact that the electronic properties of Carbon nanotubes depend strongly on their geometry \cite{Bonard:1998} points to the fact that the identification of an organizational principle or symmetry, respectively group structure,  underlying the geometry, will be an essential tool to tackle the numerous open questions in the field of Carbon onions and Carbon nanotubes. 




\end{document}